\documentclass[runningheads]{llncs}

\usepackage[T1]{fontenc}
\usepackage{graphicx}
\usepackage{amsmath,amssymb}
\usepackage{booktabs}
\usepackage{color}
\usepackage{subcaption}
\usepackage{algorithm}
\usepackage{algorithmic}
\usepackage{cite}
\usepackage{float}
\usepackage{listings}
\usepackage{xurl}
\usepackage[hidelinks]{hyperref}

\begin{document}

\title{Unlocking Python's Cores: Hardware Usage and Energy Implications of Removing the GIL}
\titlerunning{Hardware Usage and Energy Implications of Removing the GIL}

\author{José Daniel Montoya Salazar \orcidID{0009-0003-5902-7312}\\ 
\email{montoyasalazarjosedaniel@gmail.com}}

\institute{Independent Researcher}

\maketitle

\begin{abstract}
\textit{Python}'s Global Interpreter Lock (GIL) prevents \textit{Python} bytecode from executing on more than one CPU core at the same time, even when multiple threads are used. However, starting with \textit{Python}~3.13 an experimental build allows disabling the GIL. While prior work has examined speedup implications of this disabling, the effects on energy consumption and hardware utilization have received less attention. This study measures execution time, CPU utilization, memory usage, and energy consumption using four workload categories: NumPy-based computation, sequential kernels, threaded numerical workloads, and threaded object workloads, comparing GIL-enabled and free-threaded builds of \textit{Python}~3.14.2.

The results highlight a trade-off. For parallelizable workloads operating on independent data, the free-threaded build reduces execution time by up to 4$\times$, with a proportional reduction in energy consumption, and effective multi-core utilization, at the cost of an increase in memory usage. In contrast, sequential workloads do not benefit from removing the GIL and instead show a 13--43\% increase in energy consumption. Similarly, workloads where threads frequently access and modify the same objects show reduced improvements or even degradation due to lock contention. Across all workloads, energy consumption is proportional to execution time, indicating that disabling the GIL does not significantly affect power consumption, even when CPU utilization increases. When it comes to memory, the no-GIL build shows a general increase, more visible in virtual memory than in physical memory. This increase is primarily attributed to per-object locking, additional thread-safety mechanisms in the runtime, and the adoption of a new memory allocator.

These findings suggest that \textit{Python}'s no-GIL build is not a universal improvement. Developers should evaluate whether their workload can effectively benefit from parallel execution before adoption.
\end{abstract}

\section{Introduction}
\label{sec:introduction}

For several years, the electricity demand of data centers remained relatively stable due to efficiency improvements. However, according to projections by the \textit{Institut français des relations internationales} (IFRI), a French international relations think tank \cite{ENERGY_SOFTWARE}, this trend reversed after 2019 as cloud computing and artificial intelligence (AI) expanded rapidly. The information and communication technologies sector currently accounts for approximately 9\% of global electricity consumption, with data centers responsible for about 1–1.3\% of this total. In the United States, data centers could account for up to 13\% of total electricity consumption by 2030, compared to about 4\% in 2024. While in Europe AI-related computing alone could represent 4–5\% of total electricity demand \cite{ENERGY_SOFTWARE}. These projections indicate that software infrastructure is becoming a significant contributor to global energy demand.

Programming languages influence this footprint by determining how computational resources are used. \textit{Python}, the most widely used programming language \cite{PROGRAMMING_LANGUAGES_RANKING, PROGRAMMING_LANGUAGES_RANKING_BACK_UP}, prioritizes developer productivity over execution efficiency compared to lower-level languages. Pereira et al.~\cite{PEREIRA_ENERGY_EFFICIENCY_LANGUAGES} report that \textit{Python} programs can consume considerably more energy than, for example, equivalent \textit{C} implementations, with observed differences close to $76\times$. In their study, \textit{Python} ranked second-to-last in energy efficiency among the considered languages, which included \textit{C++}, \textit{Rust}, \textit{Java}, and others. Kempen et al.~\cite{ENERGY_EFFICIENCY_LANGUAGES_DEBUNK} challenges the causal interpretation and magnitude of this result. This study argues that the observed energy difference arises mainly from differences in execution time, runtime behavior, and experimental setup rather than intrinsic properties of the language itself. While these arguments are correct and well supported, they prove that if \textit{Python} ran faster, it would naturally consume less energy. This raises the question of whether \textit{Python} can actually run faster in practical environments.

In some domains, like AI, energy-intensive operations are already mitigated by design choices. Popular \textit{Python} libraries in this field use \textit{Python} primarily as a high-level interface and delegate computationally intensive tasks to \textit{C}, \textit{C++}, or \textit{Rust} code, often accelerated by GPUs. This approach, also highlighted by Kempen et al.~\cite{ENERGY_EFFICIENCY_LANGUAGES_DEBUNK}, reduces the impact of \textit{Python}’s runtime overhead in those workloads. However, AI is not the primary use case for \textit{Python}. According to the 2024 Python Developers Survey~\cite{PYTHON_DEVSURVEY2024}, developers who use \textit{Python} as their main language work across a wide range of domains. Measured as the percentage of main-\textit{Python} developers reporting use in each category, these include data analysis (49\%), web development (48\%), machine learning (42\%), data engineering (33\%), web scraping (28\%), and academic research (28\%). Some of these are non–extension-dominated fields.

In these areas, performance is determined by \textit{Python}’s own interpreter which has the Global Interpreter Lock (GIL) as a key architectural piece. It prevents multiple threads from executing \textit{Python} bytecode concurrently within a single process, limiting effective utilization of multi-core CPUs. Starting with \textit{Python~3.13} and continuing in \textit{Python~3.14}, \textit{CPython} provides an optional free-threaded build in which the GIL is disabled, enabling parallel execution of \textit{Python} threads \cite{GIL_REMOVAL}.

In this context, and motivated by the environmental importance of improving software efficiency, this study investigates whether removing the GIL can improve hardware utilization and energy efficiency. While existing works like \cite{GIL_VS_NO_GIL_3_14, GILLESS_PERF_REPORT, GIL_VS_NO_GIL_REPO} focus on execution time comparisons between GIL-enabled and free-threaded builds, hardware utilization or energy consumption have received less attention. This study addresses that gap. In addition, it reviews the factors that influence software energy consumption to explain why removing the GIL raises expectations of improvements in this matter.

\section{Literature Review}
\label{sec:literature_review}
This section reviews the main factors that determine software energy consumption. It also provides the basis for analyzing how removing the GIL may affect \textit{Python}’s hardware usage and energy behavior.

\subsection{What makes software consume energy?}

\subsubsection{Execution time}
Pereira et al.~\cite{PEREIRA_ENERGY_EFFICIENCY_LANGUAGES} show that execution time is a dominant factor in software energy consumption, but also demonstrate that it cannot be treated as the only one. Their results challenge the intuitive assumption that “faster is always greener”, showing cases in which programs with longer execution times still achieve lower overall energy consumption. Kempen et al.~\cite{ENERGY_EFFICIENCY_LANGUAGES_DEBUNK} argue that energy efficiency is not an intrinsic property of a programming language, highlighting the influence of factors such as compiler choices, runtime systems, and hardware mechanisms including voltage and frequency scaling. When these factors are tightly controlled, their experiments show that energy consumption is directly proportional to execution time. This observation follows directly from the physical definition of energy, shown in Equation~\ref{eq:energy_equation}.
\begin{equation}
    Energy (J) = Power (W) \cdot Time (s)
    \label{eq:energy_equation}
\end{equation}

However, in real-world scenarios, the power term in Equation~\ref{eq:energy_equation} cannot be assumed constant nor fully under the developer’s control. Language implementations introduce runtime overheads: interpretation, garbage collection, or memory management, which can increase average power consumption independently of execution time. Therefore, while execution time may be sufficient for controlled experimental comparisons, it can be insufficient for evaluating the energy behavior of real software systems.

\subsubsection{CPU usage}
An observation reported by Pereira et al.~\cite{PEREIRA_ENERGY_EFFICIENCY_LANGUAGES} is that CPU-related energy accounts for the majority of the total software consumption, approximately 88\% and being the remaining portion assigned to DRAM. This indicates that CPU activity is the dominant contributor to energy usage.

\begin{figure}[htbp]
    \centering
    \includegraphics[width=0.5\linewidth]{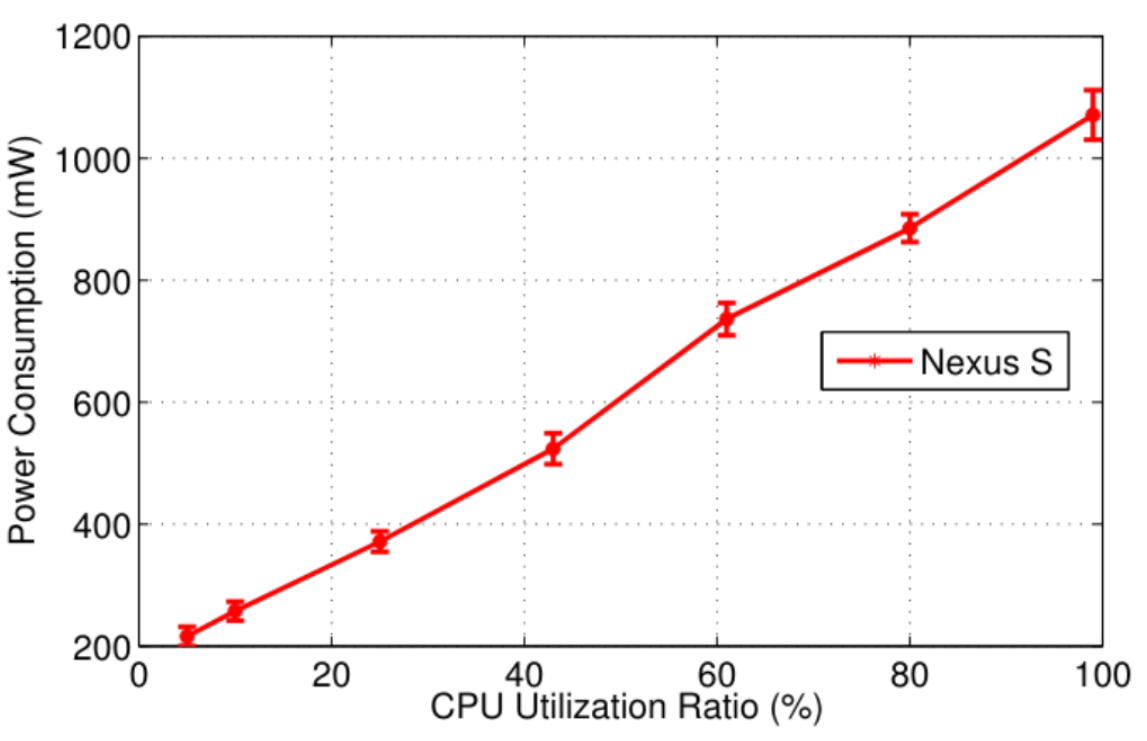}
    \caption{Energy consumption as a function of CPU usage for a mobile processor. Taken from~\cite{CPU_USAGE_ENERGY_1}.}
    \label{fig:energy_cpu_usage}
\end{figure}

This consumption increases with CPU utilization. Dargie~\cite{CPU_USAGE_ENERGY_0} and Ou et al.~\cite{CPU_USAGE_ENERGY_1} report a clear relationship between CPU usage and energy used, with approximately linear behavior for dual-core systems and quadratic behavior for single-core systems under their experimental conditions. Figure~\ref{fig:energy_cpu_usage} illustrates this relationship for a mobile processor.

And what about temperature? Since it is directly linked to energy dissipation, should it be considered an independent factor to consider? De Vogeleer et al.~\cite{DEVOGELEER_TEMPERATURE_BIAS} demonstrate that power consumption has an exponential dependence on temperature in some processors. Additionally, as temperature increases, mechanisms such as dynamic voltage or frequency scaling may be triggered, negatively affecting execution time and potentially increasing total energy consumption.

But empirical measurements confirm that CPU utilization, power dissipation, and temperature increase together under higher workloads, demonstrating a strong relationship between utilization and thermal behavior in modern processors~\cite{CPU_USAGE_ENERGY_2}. Therefore, controlling CPU usage also accounts for controlling temperature, power consumption and degradation of execution time through activation of thermal mechanisms.

\subsubsection{Memory usage}
With respect to memory, two main aspects may influence its energy consumption: peak memory usage and continuous memory access over time. Pereira et al.~\cite{PEREIRA_ENERGY_EFFICIENCY_LANGUAGES} report that there is no consistent correlation between peak memory usage and DRAM energy consumption, suggesting that memory peaks are not a reliable predictor of energy demand.

Instead, memory energy effects are primarily associated with how memory is accessed during execution. CPU caches provide the fastest access to data \cite{PHILLIPS_CPU_CACHE}, when required data is not found in cache, a miss occurs and the processor must access slower memory levels, increasing execution time. Also, if a program’s memory requirements exceed available RAM, the system may use memory swapping, where data is moved between RAM and disk storage \cite{ASHWATHNARAYANA_SWAP_MEMORY}. Since disk access is orders of magnitude slower than RAM, swapping increases execution time and energy consumption.

This is why, Kempen et al.~\cite{ENERGY_EFFICIENCY_LANGUAGES_DEBUNK} demonstrates, that memory-related energy is primarily driven by memory activity (execution time and CPU usage) rather than by memory footprint alone: a well-optimized program can have a large memory in use but excellent cache locality, keeping memory activity and execution time low.

Because memory has a smaller impact on power draw compared to CPU activity, focusing on execution time and CPU power is justified when studying energy consumption. Memory behavior is a secondary factor that influences overall energy consumption depending on access locality.

\subsubsection{Parallel processing}
Modern CPUs rely on multiple cores to exploit parallelism, which introduces both opportunities and challenges with respect to energy consumption.

Konopik et al.~\cite{PARALLEL} show in an analytical study that under idealized conditions, increasing the degree of parallelism can reduce total energy consumption by shortening execution time; even when considering the overhead required to distribute the payload among workers. While real systems may deviate from this ideal, this trade-off is well accepted by developers.

However, the decision of how many cores to use is non-trivial. Extending Equation~\ref{eq:energy_equation} and neglecting memory power contributions, total energy consumption in a multi-core system can be expressed as:
\begin{equation}
    \text{Energy} = (\text{Power}_{\text{CPU-0}} + \ldots + \text{Power}_{\text{CPU-N}}) \times \text{Time}
    \label{eq:cores_energy_equation}
\end{equation}

This formulation highlights a fundamental trade-off: increasing the number of active cores affects the overall power draw of the system, while reducing the number of active cores generally increases execution time. As noted by Kempen et al.~\cite{ENERGY_EFFICIENCY_LANGUAGES_DEBUNK}, the number of active cores is a relevant factor in determining power behavior, and variations in core utilization can therefore influence total energy consumption. In addition, idle cores are not energy-free, as leakage currents and power dissipation persist even at zero computational load \cite{PARALLEL}. 

As a result, there is no universal optimal level of parallelism and the most energy-efficient configuration requires a case-by-case evaluation. But it is clear that parallelization, through execution time reduction, offers a mechanism to reduce energy consumption. \textbf{This is the reason why the removal of GIL from \textit{Python} raises expectations of a better energy efficiency}.

\subsubsection{What makes software consume energy is}
execution time mainly, with additional influence from CPU utilization, memory access behavior, and the degree of parallel execution. Other factors, like CPU temperature and power management mechanisms affect energy consumption, but they mostly reflect CPU activity rather than independent software-level causes.

These factors are determined by programming language implementations and their runtime systems. As shown by Kempen et al.~\cite{ENERGY_EFFICIENCY_LANGUAGES_DEBUNK}, when external factors such as processor frequency and core utilization are controlled, energy consumption becomes largely proportional to execution time, and no programming language implementation is intrinsically more energy-efficient than another. However, in practical scenarios language implementations have different runtime behaviors: memory management strategies, limitations on parallel execution, etc. These influence execution time, CPU utilization, access to available hardware and, energy usage.

\subsection{Python GIL}
In the standard \textit{CPython} build, the GIL blocks multiple OS threads from executing \textit{Python} bytecode at the same time within a single process, preventing multi-threaded programs to use the cores available in the CPU. But, starting with \textit{Python~3.13}, \textit{CPython} provides an optional free-threaded build where the GIL can be disabled \cite{GIL_REMOVAL, PYTHON_FREE_THREADING_HOWTO}. This mode enables proper use of multi-core hardware, but introduces runtime overhead in comparison to the default build.

To make free-threading possible, PEP~703 \cite{GIL_REMOVAL} describes several runtime changes. The main ones will be described here to highlight possible sources of overhead and speedup from removing the GIL.

First, to operate safely without the GIL, \textit{CPython} must add synchronization and thread-safe data management around operations that were previously protected by the global lock. PEP~703 describes changes to reference counting and memory management so that object lifetime and allocation remain correct when multiple threads run simultaneously; it also requires additional coordination for cyclic garbage collection \cite{GIL_REMOVAL}. Second, to allow threads to make progress concurrently, the free-threaded build shifts from one global lock to localized mechanisms, such as per-object locks for shared containers, so that unrelated operations can proceed on different CPU cores while concurrent modifications to the same object remain safe \cite{GIL_REMOVAL}.

An important change in the free-threaded build is the use of \texttt{mimalloc} as the default memory allocator \cite{MIMALLOC_DOCS}, designed to scaling to many concurrent threads. Leijen et al.~\cite{MIMALLOC_TR} report that \texttt{mimalloc} achieves competitive or superior latency performance compared to other leading allocators. Rosner~\cite{FROSNERD_ALLOCATORS} shows that these advantages are true for small allocations mostly (below approximately $1$\,KB), which dominate typical \textit{Python} workloads.  On Linux, the free-threaded build enables eager arena reservation and commit. As mentioned in the \texttt{mimalloc} documentation \cite{MIMALLOC_DOCS}, this may reserve a large virtual memory region (on the order of $1$\,GB) for arena management. This might increase virtual memory size (VMS), but does not imply an equivalent amount of resident set size (RSS) unless the reserved memory is actually accessed.

These design choices suggest some expectations. When a workload is CPU-bound and can be divided across threads with limited sharing, disabling the GIL should increase overall CPU utilization and may reduce execution time. However, because free-threaded execution adds overhead, workloads that do not benefit from thread-level parallelism, or that heavily rely on shared objects, may see smaller speedups or even slowdowns. The free-threading HOWTO reports this overhead in the form of average slowdowns on the \texttt{pyperformance} suite, ranging from about 1\% on macOS aarch64 to about 8\% on x86-64 Linux systems \cite{PYTHON_FREE_THREADING_HOWTO}. From an energy perspective, disabling the GIL may increase instantaneous power consumption because more CPU cores can be used simultaneously. However, total energy consumption may still decrease if the resulting reduction in execution time is large enough. When parallel speedup is limited by synchronization overhead, this reduction may not occur, and total energy consumption can instead increase.

This trade-off raises the question of whether removing the GIL introduces enough CPU and memory overhead to offset any benefit in execution time given by parallelism. Because this balance is expected to be workload-dependent, it motivates the experimental comparison presented in this study.

\section{Research questions}
\label{sec:research_questions}

\begin{itemize}
    \item \textbf{RQ1:} Does the free-threaded build of \textit{Python} reduce energy consumption compared to the GIL-enabled build across different workloads and execution configurations?
    \item \textbf{RQ2:} When execution time differs between the GIL-enabled and free-threaded builds, do these differences translate into proportional changes in energy consumption?
    \item \textbf{RQ3:} How do the GIL-enabled and free-threaded builds differ in hardware resource utilization, in terms of CPU and memory usage?
\end{itemize}

\section{Methodology}
\label{sec:methodology}
This section details the measurement setup, the scenarios to be tested, and the statistical analysis used to report no-GIL/GIL ratios. All source code, profiler instrumentation, and benchmark scenarios are publicly available \cite{PROFILER_REPO}.

\subsection{Hardware and configurations}
All experiments were conducted on an x86\_64 system running Ubuntu~24.04.3~LTS with Linux kernel~6.14.0-37-generic and no GUI. The machine is equipped with an Intel Core~i7-8750H processor (6 physical cores, 12 hardware threads via hyperthreading, up to 4.1\,GHz), 16\,GiB of RAM, and 1\,TB HDD storage. The system features hybrid graphics, with an Intel UHD Graphics~630 integrated GPU and an NVIDIA GeForce GTX~1050~Ti Mobile discrete GPU. 

\textit{Python~3.14.2} was used in both configurations: the default GIL-enabled build and the free-threaded compiled with \texttt{--disable-gil}. No modifications were made to any of the implementations. For no-GIL \textit{Python} a quick test enabling $MIMALLOC\_SHOW\_STATS$ and $MIMALLOC\_VERBOSE$ confirm that $MIMALLOC\_ARENA\_EAGER\_COMMIT$ is set to $2$ by default.

\subsection{Profiler as instrument to extract data}
A custom \textit{Python} profiler was developed with a sampling-based approach to minimize measurement interference. It collects samples at an interval of 50\,ms, chosen to provide sufficient sampling frequency while limiting measurement overhead. In practice, data collection adds approximately 2\,ms per sample, yielding an effective interval of ${\sim}52$\,ms. All data collection mechanisms were designed to minimize impact on the execution of the measured program, and a test using a 10\,s $sleep$ confirmed that the profiler introduces no measurable noise overhead to the target process.\footnote{The sleep validation test's results are available at \texttt{\detokenize{results/preprocessed/sleep_*.csv}} via the link provided in Section~\ref{sec:results}.}  This is because the profiler is responsible for launching the process under analysis, then its identifier is obtained and the library \texttt{psutil} (version 7.1.3 \cite{PSUTIL_7_1_3}) is used to collect statistics associated exclusively with that process.

The metrics collected and analyzed in this study are described next:

\begin{itemize}
    \item \textbf{Execution time:} At each sampling iteration, an epoch timestamp is recorded. During post-processing, these timestamps are converted to elapsed execution time in seconds.
    \item \textbf{CPU usage:} CPU utilization is obtained from \texttt{psutil} for the target process and normalized by the number of available CPU cores. An $interval=0.0$ is used so the measurement is non-blocking.
    \item \textbf{Virtual Memory Size (VMS):} Peak amount of virtual memory reserved by the process.
    \item \textbf{Resident Set Size (RSS):} Peak amount of physical memory (RAM) currently used by the process.
    \item \textbf{Swap memory:} Peak amount of process memory swapped out to disk.
    \item \textbf{Energy consumption:} Measured using Intel RAPL via the Linux sysfs interface (\path{/sys/class/powercap/intel-rapl:*/energy_uj}), therefore it is a system-wide stat. To avoid some issues discussed by Kempen et al.\ \cite{ENERGY_EFFICIENCY_LANGUAGES_DEBUNK}, readings are kept as integers in $\mu$J, conversion to floating point is made only after subtracting consecutive readings and counter overflow is handled using \texttt{max\_energy\_range\_uj}. In addition, the profiler is designed to run without \texttt{sudo} so file permissions are configured beforehand and the energy counters can be read as an unprivileged user to avoid any extra overhead.
\end{itemize}

\subsection{Workload scenarios}
The benchmark suite designed includes native-extension workloads (\textit{Numpy}), single-threaded pure-\textit{Python} kernels, and multi-threaded programs. All scenarios follow the same structure. Inputs are generated deterministically using fixed seeds, and a 3\,s pause separates initialization from the measured regions. Setup and data-generation phases are excluded from the analysis using the \texttt{set\_tag} helper. An example is shown next.

\begin{lstlisting}[language=Python]
# Pre-build input before profiling
numbers = build_numbers(num_items)
time.sleep(3)

set_tag("start_bubble_sort")
bubble_sort(numbers)
set_tag("finish_bubble_sort")
\end{lstlisting}

\subsubsection{NumPy scenarios}
These workloads represent data-analysis patterns where \textit{Python} orchestrates optimized native code. Differences between GIL and no-GIL are expected to be small and reflect free-threading overhead or interactions with native library execution. The profiled test cases are:
\begin{itemize}
    \item \texttt{numpy\_vectorized}: arithmetic operations over vectors.
    \item \texttt{numpy\_blas}: matrix multiplication \texttt{A.dot(B)} for square matrices.
    \item \texttt{numpy\_fft}: FFT followed by magnitude reduction on signals.
\end{itemize}

\subsubsection{Sequential scenarios (single-threaded Python kernels)}
These CPU-bound programs run in a single thread and aim to measure baseline no-GIL overhead.
\begin{itemize}
    \item \texttt{mandelbrot}: nested-loop Mandelbrot over a precomputed grid.
    \item \texttt{bubble\_sort}: in-place bubble sort of a deterministic descending list.
    \item \texttt{prime\_sieve}: Sieve of Eratosthenes up to varying limits.
\end{itemize}

\subsubsection{Threaded numerical scenarios}
These scenarios use \texttt{ThreadPoolExecutor} and vary the number of workers.
\begin{itemize}
    \item \texttt{factorial}: compute factorials for inputs 0..10000.
    \item \texttt{matmul}: pure-\textit{Python} matrix multiplication over row blocks, with a fixed matrix size of $768 \times 768$.
    \item \texttt{nbody}: N-body simulation, \texttt{num\_particles=2000} and \texttt{num\_steps=10}.
\end{itemize}

\subsubsection{Threaded objects scenarios}
These workloads stress allocation, reference counting, and container operations, including shared mutable state. They are relevant because object-heavy workloads may scale differently than numerical kernels.
\begin{itemize}
    \item \texttt{json\_parse}: parse and operates \texttt{2{,}000{,}000} payloads with \texttt{json.loads}.
    \item \texttt{object\_lists\allowbreak\_nocopy} (shared mutation): uppercase values of a shared list in-place for \texttt{55{,}000{,}000} records.
    \item \texttt{object\_lists\_copy} (reduced shared mutation): uppercase values using per-thread \textit{C}-level slice copies and lists comprehension for \texttt{55{,}000{,}000} records.
    \item \texttt{object\_lists}: dataclass-based object/list transformations with \texttt{8{,}000{,}000} records.
\end{itemize}

\subsection{Repetitions and execution counts}
Each parameter point (e.g., a matrix size or worker count) is executed \textbf{ten times} under each \textit{Python} build, yielding $n=10$ noGIL/GIL comparisons per parameter point. Each no-GIL run is paired with the corresponding GIL run in post-processing.

A 60\,s cooldown is added between consecutive executions to reduce carry-over effects.

\subsection{Interpretation of results}
This section describes the mathematical analysis followed to compare and summarize the profiling results.

\subsubsection{Per-run effect size}
For each run, metrics are compared between the GIL-enabled and GIL-disabled configurations using a ratio computed on matched runs. For a metric \(X\) (e.g., execution time, CPU usage, or energy consumption), the per-run effect size is defined as:
\begin{equation}
R = \frac{X_{\text{noGIL}}}{X_{\text{GIL}}}
\end{equation}

\subsubsection{Aggregation across runs}
Because ratios represent multiplicative effects, they are aggregated using log transformation and the geometric mean. For run \(i\), the log-ratio is defined as:
\begin{equation}
L_i = \ln(R_i)
\end{equation}

The mean log-ratio across \(n\) runs is:
\begin{equation}
\bar{L} = \frac{1}{n} \sum_{i=1}^{n} L_i
\end{equation}

The aggregated effect size is obtained by exponentiating the mean log-ratio:
\begin{equation}
R_{\text{geo}} = \exp(\bar{L})
\end{equation}

\subsubsection{Confidence intervals}
Run-to-run variability is quantified in log space. Using \(s_L\) as the sample standard deviation of the log-ratios, the standard error of the mean is computed as:
\begin{equation}
SE = \frac{s_L}{\sqrt{n}}
\end{equation}

A two-sided 95\% confidence interval for the mean log-ratio is constructed using Student’s \(t\) distribution with \(n-1\) degrees of freedom:
\begin{equation}
\bar{L} \pm t_{0.975,\,n-1} \cdot SE
\end{equation}

Exponentiating the interval bounds yields a confidence interval in ratio space.

\subsubsection{Interpretation}
A confidence interval entirely below (above) 1 indicates that no-GIL is faster (slower) or uses fewer (more) resources than GIL. When the confidence interval contains 1, the observed difference is not statistically distinguishable at the 95\% confidence level.

\section{Results}
\label{sec:results}
This section presents tables summarizing the analysis just described. For brevity, only selected cases are shown, and it is safe to assume that the omitted ones follow the same trends as those reported. The complete set of results is available online at: \texttt{https://drive.google.com/drive/folders/1PDxkxgaz8Fz0YLwI}\\\texttt{XghI0AnufRDyTDs5}. Workloads were chosen to ensure that no swapping is required, avoiding the extra variable of disk access.

\begin{table}[H]
\caption{Numpy Vectorized.}
\label{tab:numpy_vectorized}
\centering
\small
\setlength{\tabcolsep}{3.5pt}
\renewcommand{\arraystretch}{1.5}

\resizebox{\columnwidth}{!}{%
\begin{tabular}{r rr rr rr rr rr}
\hline
& \multicolumn{2}{c}{Time} & \multicolumn{2}{c}{CPU} & \multicolumn{2}{c}{Energy} & \multicolumn{2}{c}{VMS} & \multicolumn{2}{c}{RAM} \\
{\scriptsize size} & $R$ & CI & $R$ & CI & $R$ & CI & $R$ & CI & $R$ & CI \\
\hline
{\scriptsize 150M} & 1.013 & 1.001--1.024 & 0.994 & 0.978--1.010 & 1.028 & 1.017--1.040 & 1.135 & 1.135--1.135 & 1.001 & 1.001--1.001 \\
{\scriptsize 170M} & 0.997 & 0.988--1.007 & 0.999 & 0.991--1.008 & 1.008 & 0.994--1.021 & 1.121 & 1.121--1.121 & 1.001 & 1.001--1.001 \\
{\scriptsize 190M} & 1.004 & 0.995--1.013 & 0.993 & 0.984--1.003 & 1.006 & 0.992--1.020 & 1.109 & 1.109--1.109 & 1.001 & 1.001--1.001 \\
{\scriptsize 210M} & 1.005 & 0.997--1.013 & 1.003 & 0.998--1.008 & 1.006 & 0.993--1.018 & 1.099 & 1.099--1.099 & 1.001 & 1.001--1.001 \\
{\scriptsize 230M} & 0.999 & 0.988--1.011 & 1.004 & 0.998--1.010 & 0.992 & 0.973--1.010 & 1.091 & 1.091--1.091 & 1.001 & 1.001--1.001 \\
{\scriptsize 250M} & 0.997 & 0.989--1.006 & 1.008 & 1.000--1.017 & 0.997 & 0.987--1.008 & 1.084 & 1.084--1.084 & 1.001 & 1.001--1.001 \\
{\scriptsize 275M} & 1.006 & 0.993--1.019 & 1.005 & 1.000--1.009 & 1.004 & 0.989--1.020 & 1.077 & 1.077--1.077 & 1.001 & 1.001--1.001 \\
\hline
\end{tabular}%
}
\end{table}

\begin{table}[H]
\caption{Sequential Bubble Sort.}
\label{tab:bubble_sort}
\centering
\small
\setlength{\tabcolsep}{3.5pt}
\renewcommand{\arraystretch}{1.5}

\resizebox{\columnwidth}{!}{%
\begin{tabular}{r rr rr rr rr rr}
\hline
& \multicolumn{2}{c}{Time} & \multicolumn{2}{c}{CPU} & \multicolumn{2}{c}{Energy} & \multicolumn{2}{c}{VMS} & \multicolumn{2}{c}{RAM} \\
{\scriptsize size} & $R$ & CI & $R$ & CI & $R$ & CI & $R$ & CI & $R$ & CI \\
\hline
{\scriptsize 5k}  & 1.334 & 1.316--1.351 & 1.010 & 1.007--1.014 & 1.346 & 1.324--1.368 & 40.332 & 40.332--40.332 & 1.232 & 1.231--1.233 \\
{\scriptsize 8k}  & 1.329 & 1.313--1.345 & 1.005 & 1.003--1.007 & 1.324 & 1.308--1.341 & 40.332 & 40.332--40.332 & 1.235 & 1.233--1.236 \\
{\scriptsize 11k} & 1.330 & 1.316--1.345 & 1.002 & 1.001--1.002 & 1.329 & 1.310--1.349 & 40.332 & 40.332--40.332 & 1.238 & 1.236--1.239 \\
{\scriptsize 14k} & 1.342 & 1.322--1.363 & 1.001 & 1.001--1.002 & 1.344 & 1.323--1.365 & 40.332 & 40.332--40.332 & 1.239 & 1.238--1.241 \\
{\scriptsize 17k} & 1.351 & 1.328--1.375 & 1.001 & 1.000--1.001 & 1.353 & 1.328--1.379 & 40.332 & 40.332--40.332 & 1.240 & 1.239--1.242 \\
{\scriptsize 21k} & 1.335 & 1.322--1.349 & 1.001 & 1.000--1.001 & 1.335 & 1.319--1.351 & 40.179 & 39.835--40.526 & 1.242 & 1.241--1.244 \\
{\scriptsize 25k} & 1.352 & 1.330--1.374 & 1.000 & 1.000--1.001 & 1.354 & 1.334--1.375 & 38.827 & 38.827--38.827 & 1.242 & 1.241--1.243 \\
\hline
\end{tabular}%
}
\end{table}

\begin{table}[H]
\caption{Sequential Prime Sieve.}
\label{tab:prime_sieve}
\centering
\small
\setlength{\tabcolsep}{3.5pt}
\renewcommand{\arraystretch}{1.5}

\resizebox{\columnwidth}{!}{%
\begin{tabular}{r rr rr rr rr rr}
\hline
& \multicolumn{2}{c}{Time} & \multicolumn{2}{c}{CPU} & \multicolumn{2}{c}{Energy} & \multicolumn{2}{c}{VMS} & \multicolumn{2}{c}{RAM} \\
{\scriptsize limit} & $R$ & CI & $R$ & CI & $R$ & CI & $R$ & CI & $R$ & CI \\
\hline
{\scriptsize 16M} & 1.167 & 1.138--1.198 & 1.004 & 1.000--1.008 & 1.170 & 1.139--1.202 & 5.558 & 5.537--5.578 & 1.186 & 1.181--1.192 \\
{\scriptsize 20M} & 1.173 & 1.139--1.209 & 1.005 & 1.000--1.009 & 1.174 & 1.141--1.208 & 4.576 & 4.570--4.582 & 1.123 & 1.111--1.136 \\
{\scriptsize 24M} & 1.134 & 1.118--1.151 & 1.003 & 1.000--1.007 & 1.139 & 1.123--1.155 & 3.885 & 3.876--3.895 & 1.128 & 1.109--1.147 \\
{\scriptsize 28M} & 1.145 & 1.117--1.174 & 1.003 & 0.999--1.006 & 1.144 & 1.118--1.171 & 3.380 & 3.377--3.383 & 1.151 & 1.149--1.153 \\
{\scriptsize 32M} & 1.136 & 1.111--1.162 & 1.003 & 1.000--1.005 & 1.140 & 1.117--1.164 & 2.997 & 2.995--2.999 & 1.146 & 1.144--1.148 \\
{\scriptsize 36M} & 1.132 & 1.122--1.143 & 1.006 & 1.003--1.008 & 1.138 & 1.125--1.150 & 2.692 & 2.689--2.694 & 1.172 & 1.171--1.173 \\
{\scriptsize 40M} & 1.146 & 1.129--1.164 & 1.004 & 1.001--1.006 & 1.151 & 1.132--1.171 & 2.439 & 2.437--2.440 & 1.146 & 1.137--1.154 \\
\hline
\end{tabular}%
}
\end{table}

\begin{table}[H]
\caption{Threaded Numerical Factorial.}
\label{tab:numerical_factorial}
\centering
\small
\setlength{\tabcolsep}{3.5pt}
\renewcommand{\arraystretch}{1.5}

\resizebox{\columnwidth}{!}{%
\begin{tabular}{r rr rr rr rr rr}
\hline
& \multicolumn{2}{c}{Time} & \multicolumn{2}{c}{CPU} & \multicolumn{2}{c}{Energy} & \multicolumn{2}{c}{VMS} & \multicolumn{2}{c}{RAM} \\
{\scriptsize w} & $R$ & CI & $R$ & CI & $R$ & CI & $R$ & CI & $R$ & CI \\
\hline
{\scriptsize 1}  & 1.212 & 1.209--1.216 & 1.050 & 1.048--1.052 & 1.215 & 1.212--1.218 & 6.053 & 6.053--6.053 & 1.058 & 1.057--1.058 \\
{\scriptsize 2}  & 0.610 & 0.600--0.619 & 2.062 & 2.058--2.067 & 0.614 & 0.605--0.624 & 6.178 & 6.178--6.178 & 1.018 & 1.016--1.020 \\
{\scriptsize 4}  & 0.307 & 0.303--0.310 & 4.050 & 4.032--4.069 & 0.307 & 0.303--0.311 & 3.958 & 3.958--3.959 & 0.953 & 0.947--0.958 \\
{\scriptsize 6}  & 0.246 & 0.244--0.248 & 6.001 & 5.940--6.064 & 0.250 & 0.247--0.253 & 3.073 & 3.070--3.076 & 0.907 & 0.884--0.932 \\
{\scriptsize 8}  & 0.247 & 0.245--0.250 & 7.913 & 7.855--7.970 & 0.251 & 0.247--0.255 & 2.596 & 2.595--2.596 & 0.885 & 0.842--0.930 \\
{\scriptsize 12} & 0.249 & 0.247--0.250 & 11.309 & 11.163--11.457 & 0.254 & 0.253--0.255 & 2.091 & 2.091--2.092 & 0.899 & 0.841--0.962 \\
\hline
\end{tabular}%
}
\end{table}

\begin{table}[H]
\caption{Threaded Numerical N-Body.}
\label{tab:numerical_nbody}
\centering
\small
\setlength{\tabcolsep}{3.5pt}
\renewcommand{\arraystretch}{1.5}

\resizebox{\columnwidth}{!}{%
\begin{tabular}{r rr rr rr rr rr}
\hline
& \multicolumn{2}{c}{Time} & \multicolumn{2}{c}{CPU} & \multicolumn{2}{c}{Energy} & \multicolumn{2}{c}{VMS} & \multicolumn{2}{c}{RAM} \\
{\scriptsize w} & $R$ & CI & $R$ & CI & $R$ & CI & $R$ & CI & $R$ & CI \\
\hline
{\scriptsize 1}  & 1.174 & 1.167--1.180 & 1.001 & 1.000--1.002 & 1.174 & 1.168--1.181 & 11.131 & 11.131--11.131 & 1.293 & 1.291--1.294 \\
{\scriptsize 2}  & 0.615 & 0.608--0.622 & 1.983 & 1.979--1.988 & 0.623 & 0.616--0.631 & 6.889 & 6.889--6.889 & 1.305 & 1.303--1.308 \\
{\scriptsize 4}  & 0.316 & 0.313--0.319 & 3.921 & 3.909--3.933 & 0.319 & 0.316--0.322 & 4.205 & 4.205--4.205 & 1.339 & 1.335--1.344 \\
{\scriptsize 6}  & 0.231 & 0.229--0.233 & 5.783 & 5.760--5.805 & 0.232 & 0.230--0.235 & 3.257 & 3.134--3.385 & 1.347 & 1.343--1.351 \\
{\scriptsize 8}  & 0.251 & 0.248--0.254 & 6.580 & 6.564--6.597 & 0.253 & 0.250--0.256 & 2.781 & 2.661--2.905 & 1.372 & 1.369--1.374 \\
{\scriptsize 12} & 0.223 & 0.222--0.224 & 11.410 & 11.371--11.448 & 0.226 & 0.224--0.227 & 2.286 & 2.198--2.377 & 1.428 & 1.425--1.430 \\
\hline
\end{tabular}%
}
\end{table}

\begin{table}[H]
\caption{Threaded Objects JSON Parse.}
\label{tab:objects_json_parse}
\centering
\small
\setlength{\tabcolsep}{3.5pt}
\renewcommand{\arraystretch}{1.5}

\resizebox{\columnwidth}{!}{%
\begin{tabular}{r rr rr rr rr rr}
\hline
& \multicolumn{2}{c}{Time} & \multicolumn{2}{c}{CPU} & \multicolumn{2}{c}{Energy} & \multicolumn{2}{c}{VMS} & \multicolumn{2}{c}{RAM} \\
{\scriptsize w} & $R$ & CI & $R$ & CI & $R$ & CI & $R$ & CI & $R$ & CI \\
\hline
{\scriptsize 1}  & 1.183 & 1.171--1.194 & 1.000 & 0.998--1.001 & 1.186 & 1.175--1.197 & 1.873 & 1.871--1.875 & 1.077 & 1.076--1.077 \\
{\scriptsize 2}  & 0.607 & 0.599--0.615 & 1.983 & 1.977--1.988 & 0.615 & 0.607--0.624 & 1.778 & 1.776--1.780 & 1.077 & 1.077--1.077 \\
{\scriptsize 4}  & 0.314 & 0.312--0.316 & 3.898 & 3.885--3.911 & 0.317 & 0.315--0.318 & 1.640 & 1.638--1.641 & 1.078 & 1.078--1.078 \\
{\scriptsize 6}  & 0.274 & 0.271--0.277 & 5.438 & 5.409--5.468 & 0.277 & 0.274--0.281 & 1.543 & 1.542--1.544 & 1.079 & 1.079--1.079 \\
{\scriptsize 8}  & 0.268 & 0.261--0.274 & 6.634 & 6.605--6.663 & 0.271 & 0.264--0.278 & 1.472 & 1.471--1.473 & 1.080 & 1.080--1.080 \\
{\scriptsize 12} & 0.275 & 0.272--0.279 & 10.912 & 10.839--10.986 & 0.279 & 0.274--0.284 & 1.374 & 1.373--1.375 & 1.083 & 1.083--1.083 \\
\hline
\end{tabular}%
}
\end{table}

\begin{table}[H]
\caption{Threaded Objects Object Lists Copy.}
\label{tab:objects_lists_copy}
\centering
\small
\setlength{\tabcolsep}{3.5pt}
\renewcommand{\arraystretch}{1.5}

\resizebox{\columnwidth}{!}{%
\begin{tabular}{r rr rr rr rr rr}
\hline
& \multicolumn{2}{c}{Time} & \multicolumn{2}{c}{CPU} & \multicolumn{2}{c}{Energy} & \multicolumn{2}{c}{VMS} & \multicolumn{2}{c}{RAM} \\
{\scriptsize w} & $R$ & CI & $R$ & CI & $R$ & CI & $R$ & CI & $R$ & CI \\
\hline
{\scriptsize 1}  & 0.875 & 0.863--0.888 & 1.714 & 1.706--1.721 & 0.879 & 0.866--0.891 & 1.197 & 1.197--1.197 & 1.147 & 1.147--1.148 \\
{\scriptsize 2}  & 0.521 & 0.518--0.524 & 2.901 & 2.885--2.917 & 0.526 & 0.519--0.532 & 1.216 & 1.178--1.254 & 1.151 & 1.148--1.155 \\
{\scriptsize 4}  & 0.355 & 0.351--0.359 & 4.294 & 4.255--4.334 & 0.360 & 0.355--0.366 & 1.383 & 1.277--1.498 & 1.169 & 1.164--1.175 \\
{\scriptsize 6}  & 0.320 & 0.315--0.325 & 5.044 & 4.958--5.132 & 0.322 & 0.317--0.326 & 1.567 & 1.418--1.731 & 1.168 & 1.161--1.175 \\
{\scriptsize 8}  & 0.301 & 0.297--0.305 & 5.938 & 5.852--6.025 & 0.300 & 0.295--0.305 & 1.452 & 1.306--1.615 & 1.163 & 1.160--1.167 \\
{\scriptsize 12} & 0.292 & 0.288--0.296 & 7.104 & 7.018--7.190 & 0.296 & 0.290--0.302 & 1.652 & 1.412--1.932 & 1.169 & 1.165--1.173 \\
\hline
\end{tabular}%
}
\end{table}

\begin{table}[H]
\caption{Threaded Objects Object Lists No Copy.}
\label{tab:objects_lists_nocopy}
\centering
\small
\setlength{\tabcolsep}{3.5pt}
\renewcommand{\arraystretch}{1.5}

\resizebox{\columnwidth}{!}{%
\begin{tabular}{r rr rr rr rr rr}
\hline
& \multicolumn{2}{c}{Time} & \multicolumn{2}{c}{CPU} & \multicolumn{2}{c}{Energy} & \multicolumn{2}{c}{VMS} & \multicolumn{2}{c}{RAM} \\
{\scriptsize w} & $R$ & CI & $R$ & CI & $R$ & CI & $R$ & CI & $R$ & CI \\
\hline
{\scriptsize 1}  & 5.222 & 5.142--5.303 & 0.476 & 0.473--0.478 & 5.269 & 5.198--5.340 & 2.256 & 2.255--2.258 & 2.308 & 2.307--2.309 \\
{\scriptsize 2}  & 4.845 & 4.618--5.084 & 0.906 & 0.886--0.926 & 4.958 & 4.715--5.213 & 2.584 & 2.574--2.595 & 2.296 & 2.286--2.307 \\
{\scriptsize 4}  & 4.519 & 4.239--4.819 & 1.761 & 1.727--1.796 & 4.622 & 4.341--4.921 & 2.574 & 2.496--2.654 & 2.278 & 2.266--2.290 \\
{\scriptsize 6}  & 4.817 & 4.610--5.032 & 2.624 & 2.584--2.665 & 4.886 & 4.682--5.098 & 2.518 & 2.497--2.540 & 2.282 & 2.267--2.297 \\
{\scriptsize 8}  & 7.889 & 7.574--8.216 & 3.592 & 3.522--3.663 & 7.967 & 7.640--8.308 & 2.669 & 2.485--2.868 & 2.254 & 2.234--2.273 \\
{\scriptsize 12} & 12.184 & 11.876--12.501 & 5.003 & 4.952--5.055 & 12.308 & 11.970--12.657 & 2.474 & 2.442--2.506 & 2.308 & 2.308--2.309 \\
\hline
\end{tabular}%
}
\end{table}

\begin{table}[H]
\caption{Threaded Objects Object Lists.}
\label{tab:objects_lists}
\centering
\small
\setlength{\tabcolsep}{3.5pt}
\renewcommand{\arraystretch}{1.5}

\resizebox{\columnwidth}{!}{%
\begin{tabular}{r rr rr rr rr rr}
\hline
& \multicolumn{2}{c}{Time} & \multicolumn{2}{c}{CPU} & \multicolumn{2}{c}{Energy} & \multicolumn{2}{c}{VMS} & \multicolumn{2}{c}{RAM} \\
{\scriptsize w} & $R$ & CI & $R$ & CI & $R$ & CI & $R$ & CI & $R$ & CI \\
\hline
{\scriptsize 1}  & 1.277 & 1.269--1.285 & 1.207 & 1.200--1.213 & 1.281 & 1.271--1.292 & 1.284 & 1.283--1.285 & 1.150 & 1.150--1.150 \\
{\scriptsize 2}  & 0.810 & 0.804--0.817 & 2.114 & 2.101--2.127 & 0.812 & 0.803--0.821 & 1.364 & 1.291--1.441 & 1.174 & 1.167--1.180 \\
{\scriptsize 4}  & 0.565 & 0.556--0.575 & 3.457 & 3.432--3.482 & 0.568 & 0.559--0.578 & 1.506 & 1.363--1.664 & 1.300 & 1.259--1.343 \\
{\scriptsize 6}  & 0.545 & 0.535--0.555 & 4.062 & 3.986--4.139 & 0.550 & 0.541--0.561 & 1.593 & 1.495--1.698 & 1.368 & 1.335--1.401 \\
{\scriptsize 8}  & 0.514 & 0.505--0.523 & 5.090 & 5.023--5.158 & 0.517 & 0.507--0.527 & 1.631 & 1.538--1.730 & 1.364 & 1.321--1.409 \\
{\scriptsize 12} & 0.563 & 0.553--0.573 & 6.302 & 6.205--6.399 & 0.569 & 0.557--0.581 & 1.795 & 1.653--1.949 & 1.627 & 1.578--1.678 \\
\hline
\end{tabular}%
}
\end{table}

\subsection{Discussion}
\label{subsec:discussion}

This section analyzes the results of each set of scenarios in order to answer the research questions.

\subsection{NumPy Scenarios}
For the NumPy scenarios, behavior was consistent across all three benchmarks. Disabling the GIL yields no meaningful speedup. Across 21 parameter points, all execution-time ratios fall within $0.997$--$1.015$, and 20 of the 21 confidence intervals include the neutral value of~1.0. The remaining case shows a negligible slowdown of $1.3\%$, corresponding to ${\sim}86$\,ms difference. This was expected and exemplifies the approach highlighted by Kempen et al.~\cite{ENERGY_EFFICIENCY_LANGUAGES_DEBUNK}: using \textit{Python} as a high-level orchestrator while delegating intensive work to native libraries neutralizes \textit{Python}'s runtime overhead.

CPU utilization ratios are similarly neutral, confirming that both builds exercise this resource identically. Also, the pre-processed data shows that all the cores of the CPU are being used for both implementations, indicating that Numpy successfully releases the GIL in the default build.

\textbf{RQ1 and RQ2:} Energy consumption mirrors execution time almost exactly. All energy ratios fall within $0.992$--$1.028$, with confidence intervals that tend to contain~1.0. Because both execution time and CPU utilization are unchanged energy differences are negligible.

\textbf{RQ3:} The only notable hardware difference is in virtual memory size (VMS). The no-GIL build shows VMS ratios of $1.070$--$1.714$ (i.e., $7.0\%$--$71.4\%$ higher). In absolute terms, this corresponds to approximately $1$\,GB of additional virtual space. However, resident memory (RSS) ratios remain close to $1.0$, indicating that the additional virtual memory is not sustained by physical RAM in these scenarios. This reflects the memory behavior expected from $mimalloc$, visible here as a baseline since Numpy uses standard \textit{C} allocation mechanisms.

In summary, for workloads dominated by native extensions that release the GIL, removing the interpreter lock has no practical effect on execution time, CPU utilization, or energy consumption. The free-threaded build does introduce considerable virtual memory overhead but otherwise behaves equivalently to the GIL-enabled interpreter.

\subsection{Sequential Scenarios}
For single-threaded workloads, removing the GIL consistently degrades performance. Bubble sort exhibits execution-time ratios of $1.329$--$1.352$ across all input sizes, indicating that the no-GIL build is $33$--$35\%$ slower than the GIL-enabled interpreter. Mandelbrot shows the same trend, with slowdowns between $40$--$43\%$. The prime sieve exhibits a smaller but still significant overhead, with ratios of $1.132$--$1.173$ ($13$--$17\%$ slower). All confidence intervals lie entirely above~1.0, confirming that these slowdowns are statistically significant. This overhead is expected, although it is larger than the slowdowns reported by the \textit{Python} developers HOWTO~\cite{PYTHON_FREE_THREADING_HOWTO}, that showed average penalties of $8\%$.

CPU utilization ratios are essentially neutral, confirming that both builds use a single core identically. This is expected as without multi-threading the workload remains bound to one CPU core.

\textbf{Answering RQ1 and RQ2:} Energy consumption tracks execution time with high fidelity. Because CPU utilization is unchanged, power draw remains constant between builds, and energy differences are determined entirely by execution time. This provides further empirical support for the finding of Kempen et al.~\cite{ENERGY_EFFICIENCY_LANGUAGES_DEBUNK} that energy is proportional to execution time when power behavior is held constant. Thus, for sequential workloads, no-GIL increases energy consumption by the same factor as it increases execution time.

\textbf{Answering RQ3:} Memory overhead is substantial. Virtual memory ratios for bubble sort are approximately $40\times$ higher in the no-GIL build, while prime sieve exhibits ratios of $2.4$--$5.5\times$. This difference is largely explained by workload scale: as the sieve’s own memory footprint increases, it becomes comparable to the ground overhead introduced by the no-GIL runtime.

More importantly, unlike the NumPy scenarios, RAM usage also increases. Bubble sort shows RSS ratios of $1.23$--$1.24\times$, while prime sieve exhibits $1.12$--$1.18\times$. This shows the free-threaded allocation mechanisms consuming actual physical memory when execution is dominated by the \textit{Python} interpreter.

In summary, for sequential pure-\textit{Python} workloads, removing the GIL offers no benefit and imposes measurable costs: longer execution times, proportionally higher energy consumption, and increased memory usage. These results set a baseline overhead that must be overcome by parallelism for no-GIL to be advantageous.

\subsection{Threaded Numerical Scenarios}
With a single worker, the no-GIL build exhibits the expected sequential overhead: time ratios of $1.17$--$1.21$, consistent with the findings from the sequential scenarios. However, as worker count increases, no-GIL achieves substantial speedups. At 6 workers, time ratios drop to $0.23$--$0.25$ across all three benchmarks, representing ${\sim}4\times$ faster execution. The matrix multiplication shows the same trend as N-body.

CPU ratios show up to~${\sim}11\times$ more usage of this resource from the no-GIL build, indicating that threads execute concurrently across multiple cores. In contrast, the GIL-enabled build shows no improvement with additional threads, indicating that execution remains constrained to a single core.

The diminishing benefits after 6 workers are expected. The test machine has 6 physical cores (12 logical cores with hyperthreading), so scaling beyond 6 workers provides only modest improvements. This illustrates the trade-off between potential parallelization and the actual number of available cores.

\textbf{Answering RQ1 and RQ2:} Energy consumption tracks execution time with high fidelity. At 6 workers, energy ratios are $0.23$--$0.26$, nearly identical to the time ratios. This represents a ${\sim}4\times$ reduction in energy consumption. Despite activating more CPU cores (which increases power draw), the reduction in execution time dominates the energy equation. This confirms the finding of Kempen et al.~\cite{ENERGY_EFFICIENCY_LANGUAGES_DEBUNK} that aggressive parallelization is nearly always energy-efficient: on their platform, doubling cores added only ${\sim}31$W while still providing larger time and energy savings.

\textbf{Answering RQ3:} CPU utilization increases with the number of workers under no-GIL, reaching ${\sim}11\times$ at 12 workers. This confirms that the free-threaded build enables true parallel execution of \textit{Python} bytecode across multiple cores. VMS overhead is also observed in these scenarios (up to $11\times$ higher), but its relative impact decreases as the number of threads increases. This indicates that the GIL build scales worse in this aspect, as its memory usage tends to converge toward no-GIL one with increasing thread count. RSS overhead remains modest, with a largest absolute increment of $8$~MB observed in the N-body scenario with 12 threads. Interestingly, the factorial scenario shows slightly lower RSS under no-GIL after four threads. An explanation is the workload-specific behavior: this benchmark shares a single read-only input list and performs arithmetic operations inside \texttt{math.factorial}, resulting in less per-thread \textit{Python} container allocation than \texttt{matmul} and \texttt{nbody}. In addition, \texttt{mimalloc} performs well for small allocations~\cite{MIMALLOC_TR, FROSNERD_ALLOCATORS}. Therefore, this small RSS reduction should be interpreted as an effect specific to this case, not as evidence of a general no-GIL memory advantage. In any case, the VMS overhead after removing the GIL is notorious.

In summary, threaded numerical workloads are the ideal use case for the free-threaded build. When work can be partitioned across threads with low shared state, no-GIL delivers clear speedups up to the physical core count, with proportional energy savings.

\subsection{Threaded Objects Scenarios}
These results show that behavior depends critically on whether threads share mutable state or not.

\textbf{Workloads with minimal contention perform well.} The \texttt{json\_parse} and \texttt{object\_lists\_copy} benchmarks exhibit scaling comparable to the numerical scenarios. At 6--8 workers, \texttt{json\_parse} achieves time ratios of ${\sim}0.27$, and \texttt{object\_lists\_copy} reaches $0.30$--$0.32$. These workloads share a key property: threads operate on largely independent data. In \texttt{json\_parse}, threads read distinct slices of JSON payloads from a shared list but accumulate results in thread-local structures. In \texttt{object\_lists\_copy}, each thread copies its assigned slice to a local list before transforming it. Because threads do not mutate shared objects, per-object locks are rarely contended, allowing parallelism to proceed efficiently. \footnote{A plausible explanation for the absence of one-thread degradation in \texttt{object\_lists\_copy} is that it is dominated by \textit{C}-level bulk operations (slice copy and lists comprehension) on thread-local data, so lock contention is negligible. This speedup is case-specific rather than a general no-GIL advantage.}

The \texttt{object\_lists} benchmark shows moderate improvement. At 8 workers, the time ratio is $0.51$, less than the ${\sim}3$--$4\times$ speedup achieved by the previous workloads. This benchmark creates class instances and performs list concatenations within each thread's slice, introducing more allocation pressure and reference counting overhead. Still, because each thread operates on its own slice of data and builds independent output structures, contention remains manageable.

\textbf{Shared mutable state causes severe degradation.} The \texttt{object\_lists\allowbreak\_nocopy} benchmark presents clear contrast. At a single worker, no-GIL is $5.2\times$ slower. As workers increase, performance gets worse: at 12 workers, no-GIL is $12.18\times$ slower. This is the opposite of the scaling observed in all other threaded scenarios.

This workload repeatedly creates new strings and writes them into a shared list. Under PEP~703 free-threading, these operations incur additional runtime costs (object management and synchronization), and at higher worker counts, concurrent mutation of the same container can add lock contention~\cite{GIL_REMOVAL}. Together, these effects shadow any parallel benefit.

The CPU utilization data supports this interpretation. For \texttt{object\_lists\allowbreak\_nocopy}, the CPU ratio at 12 workers is only $5.0$, below the $6$--$11\times$ achieved by other benchmarks at the same worker count. At low thread counts (one and two), \texttt{object\_lists\allowbreak\_nocopy} even performs worse than GIL-enabled \textit{Python}. The large time difference (up to \(20\)s) and the lower CPU usage, indicate that execution time is dominated by memory-management overhead rather than CPU occupation.

\textbf{Answering RQ1 and RQ2:} Energy consumption diverges dramatically based on sharing patterns. For \texttt{json\_parse} and \texttt{object\_lists\_copy}, energy ratios at 6--8 workers are $0.27$--$0.32$, representing ${\sim}3\times$ energy savings, consistent with the time-energy proportion observed throughout this study. For \texttt{object\_lists\allowbreak\_nocopy}, however, no-GIL consumes over 12 times more energy than GIL for the same computation. This is the worst-case scenario for the free-threaded build: not only does parallelism fail to materialize, but the overhead of contended locking increases both execution time and energy consumption.

\textbf{Answering RQ3:} Memory behavior varies by workload. VMS is consistently higher for no-GIL cases, with a largest observed VMS ratio of $2.67\times$ in the \texttt{object\_lists\allowbreak\_nocopy} benchmark with 8 threads, corresponding to a maximum absolute difference of $12.71$~GB. RSS overhead is also the highest in this set, reaching $2.3\times$ and up to $7$~GB in magnitude difference. Overall, these results confirm that no-GIL tends to use memory more intensively, and CPU usage is generally higher for no-GIL, with the previously discussed exception of \texttt{object\_lists\allowbreak\_nocopy}.

In summary, object-heavy workloads perform well under no-GIL when threads operate on independent data. However, concurrent mutation of shared containers causes severe performance and energy degradation due to per-object lock contention. This illustrates an important trade-off in free-threaded \textit{Python}: removing the global lock enables parallelism but exposes applications to contention that the GIL previously hid. Developers adopting free-threaded \textit{Python} must structure their code to minimize shared mutable state.

\subsection{Summary of Findings}
This section synthesizes the results across all scenarios to provide consolidated answers to the research questions.

\vspace{\baselineskip}
\textbf{RQ1: Does \textit{Python} without the GIL reduce energy consumption compared to \textit{Python} with the GIL?}

The answer is: conditionally yes. Energy reduction depends entirely on whether the workload can exploit parallelism without suffering contention overhead. Table~\ref{tab:summary_rq1} summarizes the observed energy ratios across scenario categories.

\begin{table}[H]
\caption{Summary of energy consumption ratios across scenarios categories at optimal worker counts (6+).}
\label{tab:summary_rq1}
\centering
\small
\setlength{\tabcolsep}{3.5pt}
\renewcommand{\arraystretch}{1.2}
\begin{tabular}{l c c}
\hline
\textbf{Scenario Category} & \textbf{Energy Ratio} & \textbf{Interpretation} \\
\hline
Native extensions & ${\sim}1.0$ & No difference \\
Single-threaded & 1.13--1.43 & 13--43\% \textit{more} \\
Threaded numerical & 0.23--0.26 & 74--77\% \textit{less} \\
Threaded objects (low contention) & 0.27--0.55 & 45--73\% \textit{less} \\
Threaded objects (high contention) & 4.8 & 380 \% \textit{more} \\
\hline
\end{tabular}
\end{table}

\vspace{\baselineskip}
\textbf{RQ2: When execution time differs between GIL and no-GIL, do those differences translate into proportional changes in energy consumption?}

Yes. Across all 84 parameter points in this study, energy tracks time with high fidelity. The mean absolute difference between time and energy ratios, over all parameter points, is less than $1\%$. For the hardware and scenarios presented, it is safe to state that energy differences between GIL and no-GIL \textit{Python} are dominated by execution time rather than variations in power draw.

Notably, this proportionality holds even when CPU utilization differs substantially between builds. In the threaded numerical scenarios, no-GIL uses up to $11\times$ more CPU (reaching average values of up to 95\% of occupation during the program execution), which increases power draw. However, the corresponding reduction in execution time compensates, resulting in energy savings. This is consistent with Kempen et al.~\cite{ENERGY_EFFICIENCY_LANGUAGES_DEBUNK}, who show that power scales sublinearly with active cores, meaning that parallelization is almost always energy-efficient if it comes with meaningful speedups.

The practical implication is significant: \textbf{for \textit{Python} 3.14 applications, optimizing for execution time is effectively equivalent to optimizing for energy consumption.} Developers must not treat energy as a separate optimization target: reducing execution time, whether through algorithmic improvements, parallelization, or runtime selection, will yield proportional energy savings.

\vspace{\baselineskip}
\textbf{RQ3: How do GIL and no-GIL differ in hardware resource utilization?}

\begin{itemize}
    \item \textbf{CPU utilization:} Under no-GIL, CPU usage scales with thread count, reaching up to $11\times$ more utilization.
    
    \item \textbf{Virtual memory (VMS):} The no-GIL build consistently uses more virtual address space ($1.1$--$40\times$ depending on workload), reflecting per-object lock structures, thread-local storage and $mimalloc$ behavior. This overhead is proportionally smaller for memory-intensive workloads confirming a baseline memory usage by no-GIL \textit{Python}.
    
    \item \textbf{Resident memory (RSS):} Physical memory overhead is modest for most workloads ($1.0$--$1.6\times$) but elevated for contention-heavy scenarios ($2.3\times$).
\end{itemize}

\subsection{The Cost of Free Threading}
While the results demonstrate that free-threading can bring substantial energy savings for parallel workloads, an honest reflection reveals a disappointing picture for general-purpose adoption.

The sequential overhead is unavoidable, and its impact applies to \emph{all} code running under the free-threaded build, regardless of whether threads are used or not. Real-world applications are never 100\% parallel, and if sequential code costs more energy, a substantial fraction of total runtime must be spent in parallel sections (with independent data) to achieve net savings.

The memory overhead increases this concern. The VMS and RSS increment is present during the whole execution, not just during parallel phases. For memory-constrained environments like embedded systems or containerized microservices with strict limits, this overhead may be too much even if the energy trade-off is favorable.

These observations do not diminish the value of free-threading for workloads that can exploit it. The $4\times$ energy reduction observed in parallel numerical scenarios is substantial and meaningful. However, for the broader \textit{Python} ecosystem, where most applications mix sequential and parallel phases, and many rely on native extensions, the free-threaded build is unlikely to be a universal improvement.

\section{Threats to Validity}
\label{sec:threats_to_validity}

\begin{itemize}

\item \textbf{Evolving runtime implementation.}
Experiments used \textit{Python~3.14.2}. Even though this represents a production release, this free-threaded build is the first one not marked as experimental, and performance may be optimized in future releases.

\item \textbf{Choice of workload size.}
One potential concern is that the smallest workload values used in the experiments may be insufficient to reveal cases where shorter execution time leads to higher energy consumption due to increased instantaneous power. While this concern is valid, execution times considerably shorter would make CPU activity measurements challenging because of limited sampling resolution. Moreover, for very short-running programs, even large ratios would translate into very small magnitudes of absolute differences, limiting their practical relevance.

\item \textbf{Single-machine measurements.}
All experiments were conducted on a single laptop and results may not generalize to other platforms.

\item \textbf{No comparison with multiprocessing.}
This study intentionally excludes \texttt{multiprocessing}-based parallelism because threads and processes represent different programming models with different overhead profiles. The goal of this study is to evaluate whether free-threading delivers on its promise of actual shared-memory parallelism, not to compare threading against multiprocessing.

\end{itemize}

\section{Conclusion}
\label{sec:conclusion}
This study evaluated the energy consumption, execution time, and hardware utilization of \textit{Python}'s free-threaded (no-GIL) build compared to the traditional GIL-enabled build across four categories of workloads: NumPy-based computation, sequential pure-\textit{Python} kernels, threaded numerical scenarios, and threaded object scenarios.

The results reveal a clear trade-off. For CPU-bound workloads that can be effectively parallelized across independent data, the free-threaded build delivers substantial benefits: execution times reduced by up to $4\times$ and energy consumption reduced proportionally, with the best results observed at high thread counts operating on thread-local data. These gains confirm that removing the GIL enables \textit{Python} threads to utilize multi-core hardware effectively.

However, these benefits come at a cost. Sequential workloads suffer a 13--43\% energy penalty under the free-threaded build. And for workloads with shared mutable state, even threaded execution may see degraded performance due to lock contention. Additionally, the presence of an overhead in memory usage is consistent across all scenarios, which can have big ratios and absolute differences. This appears from the mechanisms required to operate safely without the GIL.

For practical adoption, developers should consider the parallel fraction of their workloads and memory available. Applications that spend a large portion of their runtime in parallelizable, independent-data computation will benefit from the free-threaded build as execution time will be reduced. Applications that are predominantly sequential or rely heavily on shared mutable containers should remain on the GIL-enabled build. And Numpy-dominated workloads will see negligible differences either way, as computation occurs outside the \textit{Python} interpreter. In general for \textit{Python} applications, optimizing for execution time is equivalent to optimizing for energy consumption.

In summary, \textit{Python}'s free-threaded build is not a universal improvement. It is a specialized tool that unlocks significant energy and performance benefits for a subset of workloads while adds overhead on others. Understanding this trade-off is essential for developers seeking to build energy-efficient \textit{Python} applications in a multi-core environment.

\bibliographystyle{splncs04}
\bibliography{references}

@misc{ENERGY_SOFTWARE,
  author       = {L. de Roucy-Rochegonde and A. Buffard},
  title        = {{AI, Data Centers and Energy Demand: Reassessing and Exploring the Trends}},
  howpublished = {{IFRI Papers}},
  month        = feb,
  year         = {2025},
  url          = {https://www.ifri.org/sites/default/files/2025-02/ifri_buffard-rochegonde_ai_data_centers_energy_2025_2.pdf},
  note         = {Accessed: Mar. 4, 2026}
}

@misc{PROGRAMMING_LANGUAGES_RANKING,
  author       = {{TIOBE Software BV}},
  title        = {{TIOBE Index for February 2026}},
  howpublished = {{TIOBE Programming Community Index}},
  month        = feb,
  year         = {2026},
  url          = {https://www.tiobe.com/tiobe-index/},
  note         = {Accessed: Mar. 4, 2026}
}

@misc{PROGRAMMING_LANGUAGES_RANKING_BACK_UP,
  author       = {D. Yigit},
  title        = {{TIOBE Index Ratings}},
  howpublished = {{GitHub repository}},
  year         = {2021},
  url          = {https://github.com/toUpperCase78/tiobe-index-ratings/blob/master/Tiobe_Index_February2026.csv},
  note         = {Accessed: Mar. 4, 2026}
}

@inproceedings{PEREIRA_ENERGY_EFFICIENCY_LANGUAGES,
  author    = {R. Pereira and M. Couto and F. Ribeiro and R. Rua and J. Cunha and J. P. Fernandes and J. Saraiva},
  title     = {{Energy efficiency across programming languages: How do energy, time, and memory relate?}},
  booktitle = {{Proceedings of the ACM SIGPLAN International Conference on Software Language Engineering (SLE)}},
  year      = {2017},
  pages     = {256--267},
  doi       = {10.1145/3136014.3136031}
}

@article{ENERGY_EFFICIENCY_LANGUAGES_DEBUNK,
  author  = {N. van Kempen and H.-J. Kwon and D. T. Nguyen and E. D. Berger},
  title   = {{It's Not Easy Being Green: On the Energy Efficiency of Programming Languages}},
  journal = {{arXiv preprint arXiv:2410.05460}},
  year    = {2024},
  url     = {https://arxiv.org/abs/2410.05460},
  note    = {Accessed: Mar. 4, 2026}
}

@misc{PYTHON_DEVSURVEY2024,
  author       = {{Python Software Foundation and JetBrains}},
  title        = {{Python Developers Survey 2024 Results}},
  month        = oct,
  year         = {2024},
  howpublished = {{Oct.--Nov. 2024}},
  url          = {https://lp.jetbrains.com/python-developers-survey-2024/},
  note         = {Accessed: Mar. 4, 2026}
}

@misc{GIL_REMOVAL,
  author       = {S. Gross},
  title        = {{PEP 703 -- Making the Global Interpreter Lock Optional in CPython}},
  howpublished = {{Python Enhancement Proposals}},
  year         = {2023},
  url          = {https://peps.python.org/pep-0703/},
  note         = {Accessed: Mar. 4, 2026}
}

@misc{GIL_VS_NO_GIL_3_14,
  author       = {B. Marchand},
  title        = {{Benchmarks of Python 3.14b2 with --disable-gil}},
  howpublished = {{DEV}},
  month        = jun,
  year         = {2025},
  url          = {https://dev.to/basilemarchand/benchmarks-of-python-314b2-with-disable-gil-1ml3},
  note         = {Accessed: Mar. 4, 2026}
}

@techreport{GILLESS_PERF_REPORT,
  author      = {F. Hennecke},
  title       = {{Impact of GIL-less CPython on Performance and Compatibility}},
  institution = {University of G\"ottingen},
  type        = {Seminar report},
  month       = apr,
  year        = {2025},
  url         = {https://hps.vi4io.org/_media/teaching/autumn_term_2024/stud/scap/frederik_hennecke.pdf}
}

@misc{GIL_VS_NO_GIL_REPO,
  author       = {V. Margot},
  title        = {{Performance Comparison: Python 3.13 vs Python 3.13t (No GIL)}},
  howpublished = {{GitHub repository}},
  year         = {2025},
  url          = {https://github.com/VMargot/gil-vs-no-gil}
}

@article{CPU_USAGE_ENERGY_0,
  author  = {W. Dargie},
  title   = {{A Stochastic Model for Estimating the Power Consumption of a Processor}},
  journal = {{IEEE Transactions on Computers}},
  volume  = {64},
  number  = {5},
  pages   = {1316--1329},
  year    = {2015},
  doi     = {10.1109/TC.2014.2315629}
}

@inproceedings{CPU_USAGE_ENERGY_1,
  author    = {Z. Ou and S. Dong and J. Dong and J. Nurminen and A. Yl{\"a}-J{\"a}{\"a}ski and R. Wang},
  title     = {{Characterizing the energy impact of concurrent network-intensive applications on mobile platforms}},
  booktitle = {{Proceedings of the ACM MobiCom}},
  pages     = {23--28},
  year      = {2013},
  doi       = {10.1145/2505906.2505909}
}

@inproceedings{DEVOGELEER_TEMPERATURE_BIAS,
  author    = {K. De Vogeleer and G. Memmi and P. Jouvelot and F. Coelho},
  title     = {{Modeling the temperature bias of power consumption for nanometer-scale CPUs in application processors}},
  booktitle = {{Proceedings of the International Conference on Embedded Computer Systems: Architectures, Modeling, and Simulation (SAMOS XIV)}},
  pages     = {172--180},
  year      = {2014},
  doi       = {10.1109/SAMOS.2014.6893209}
}

@article{CPU_USAGE_ENERGY_2,
  author  = {A. M. Haywood and J. Sherbeck and P. Phelan and G. Varsamopoulos and S. K. S. Gupta},
  title   = {{The relationship among CPU utilization, temperature, and thermal power for waste heat utilization}},
  journal = {{Energy Conversion and Management}},
  volume  = {95},
  pages   = {297--303},
  year    = {2015},
  doi     = {10.1016/j.enconman.2015.01.088}
}

@misc{PHILLIPS_CPU_CACHE,
  author       = {G. Phillips},
  title        = {{How Does CPU Cache Work and What Are L1, L2, and L3 Cache?}},
  howpublished = {{MakeUseOf}},
  month        = jan,
  year         = {2023},
  url          = {https://www.makeuseof.com/tag/what-is-cpu-cache/},
  note         = {Accessed: Mar. 4, 2026}
}

@misc{ASHWATHNARAYANA_SWAP_MEMORY,
  author       = {S. Ashwathnarayana},
  title        = {{Swap Memory -- When and How to Use It on Your Production Systems or Cloud-Provided VMs}},
  howpublished = {{Netdata Blog}},
  month        = may,
  year         = {2023},
  url          = {https://www.netdata.cloud/blog/swap-memory-when-and-how-to-use-it-on-your-production-systems-or-cloud-provided-vms/},
  note         = {Accessed: Mar. 4, 2026}
}

@article{PARALLEL,
  author  = {M. Konopik and T. Korten and E. Lutz and H. Linke},
  title   = {{Fundamental energy cost of finite-time parallelizable computing}},
  journal = {{Nature Communications}},
  volume  = {14},
  pages   = {447},
  year    = {2023},
  doi     = {10.1038/s41467-023-36020-2}
}

@misc{PYTHON_FREE_THREADING_HOWTO,
  author       = {{Python Software Foundation}},
  title        = {{Python Support for Free Threading}},
  howpublished = {{Python 3 Documentation}},
  year         = {2026},
  url          = {https://docs.python.org/3/howto/free-threading-python.html},
  note         = {Accessed: Mar. 4, 2026}
}

@misc{MIMALLOC_DOCS,
  author       = {{Microsoft Corporation}},
  title        = {{mimalloc documentation}},
  url          = {https://microsoft.github.io/mimalloc/},
  note         = {Accessed: Feb. 16, 2026}
}

@techreport{MIMALLOC_TR,
  author      = {D. Leijen and B. Zorn and L. de Moura},
  title       = {{Mimalloc: Free List Sharding in Action}},
  institution = {{Microsoft Research}},
  number      = {{MSR-TR-2019-18}},
  month       = jun,
  year        = {2019},
  url         = {https://www.microsoft.com/en-us/research/wp-content/uploads/2019/06/mimalloc-tr-v1.pdf},
  note        = {Accessed: Feb. 16, 2026}
}

@misc{FROSNERD_ALLOCATORS,
  author       = {F. Rosner},
  title        = {{libmalloc, jemalloc, tcmalloc, mimalloc -- Exploring Different Memory Allocators}},
  howpublished = {{DEV Community}},
  year         = {2023},
  url          = {https://dev.to/frosnerd/libmalloc-jemalloc-tcmalloc-mimalloc-exploring-different-memory-allocators-4lp3},
  note         = {Accessed: Feb. 16, 2026}
}

@misc{PROFILER_REPO,
  author       = {J. Montoya},
  title        = {{System Profiler}},
  howpublished = {{GitHub repository}},
  year         = {2024},
  url          = {https://github.com/Joseda8/profiler/tree/v0.1.0},
  note         = {Accessed: Mar. 4, 2026}
}

@misc{PSUTIL_7_1_3,
  author       = {G. Rodol\`a},
  title        = {{psutil 7.1.3}},
  howpublished = {{Python Package Index (PyPI)}},
  year         = {2026},
  url          = {https://pypi.org/project/psutil/7.1.3/},
  note         = {Accessed: Mar. 4, 2026}
}

\end{document}